\documentclass[superscriptaddress,
twocolumn,showpacs,preprintnumbers,amsmath,amssymb]{revtex4}
\usepackage{graphicx}% Include Figure files
\usepackage{dcolumn}% Align table columns on decimal point
\usepackage{bm}% bold math
\usepackage{latexsym}
\usepackage{graphicx}
\usepackage{psfrag}               
\usepackage{epsfig}

\begin{document}

\title{Length control of microtubules by depolymerizing motor proteins}
\author{Bindu S. Govindan\footnote{\scriptsize{E-mail: bindu@mri.ernet.in}}}
\author{Manoj Gopalakrishnan}
\affiliation{Harish-Chandra Research Institute, Allahabad 211019,
  India}
\author{Debashish Chowdhury}\affiliation{Department of Physics, Indian
    Institute of Technology, Kanpur 208016, India.}

\date{\today}

\begin{abstract}
In many intracellular processes, the length distribution of microtubules 
is controlled by depolymerizing motor proteins. 
Experiments have shown that, following non-specific binding to the surface of a 
microtubule, depolymerizers are transported to the microtubule tip(s) 
by diffusion or directed walk and, then, depolymerize the microtubule 
from the tip(s) after accumulating there. We develop a quantitative 
model to study the depolymerizing action of such a generic motor 
protein, and its possible effects on the length distribution of 
microtubules. We show that, when the motor protein concentration in 
solution exceeds a critical value, a steady state is reached where the length
distribution is, in general, non-monotonic with a single peak. However, 
for highly processive motors and large motor densities, this distribution 
effectively becomes an exponential
decay. Our findings suggest that such motor proteins may be selectively used
by the cell to ensure precise control of MT lengths. The model is also used to
analyze experimental observations of motor-induced depolymerization.
\end{abstract}

\maketitle

\section{Introduction}

In eukaryotic cells, microtubules (MT) are one type of cytoskeletal
filaments, which perform several roles: they serve as tracks for
intracellular molecular motor transport, provide structural rigidity to the
cell and assemble to form a spindle during metaphase for the purpose of
separation of duplicated chromosomes to the mother and daughter cells.
MT are highly dynamic. In {\it in vitro} situations, it is observed that 
a growing MT can suddenly start shrinking; this ``catastrophe'' 
is triggered by the loss of the GTP cap by hydrolysis of GTP 
molecules attached to the tubulin subunits of MT. 
A shrinking MT occasionally get ``rescued'' and start polymerizing 
again. This unusual process of polymerization and
depolymerization is referred to as `dynamic instability'
~\cite{desai97}.

The situation inside the living cell is much more complex and the MT kinetics
{\it in vivo} is regulated by many proteins. In particular, several motor proteins
belonging to the kinesin family are now known to function as MT
depolymerizers~\cite{hunter00,moore05,kinoshita06,howard07} and are crucial 
in the formation and maintenance of the mitotic
spindle~\cite{wordeman05}. These include XKCM1/MCAK (mitotic
centromere-associated kinesin) and Kif2A belonging to the kinesin-13 family
and Kip3 proteins belonging
to kinesin-8 family. MCAK is known to be active at kinetochores (the protein
structure which facilitates MT attachment to chromosomes), whereas Kif2A
is associated with centrosomes and Kip3p regulates microtubule-cortical
interactions. Depletion/inhibition of XKCM1/MCAK has been shown to affect
spindle length and the poleward motion of chromosomes during anaphase, while
deletion of kinesin-8 proteins leads to defects in positioning of the
spindle\cite{wordeman05}. However, the mechanism of depolymerization of MT by these
motor proteins is only incompletely understood~\cite{hunter03}.

Recent {\it in vitro} depolymerization experiments with surface-immobilized MT
have demonstrated that the depolymerizing kinesins MCAK~\cite{howard06} and 
Kip3p~\cite{varga06,gupta06}use a
`reduction of dimensionality'
mechanism to target the MT tips for depolymerization: after binding
non-specifically to the MT surface, these proteins use diffusion (MCAK) or
directed walk (Kip3p) to reach the tip(s), and their accumulation induces 
depolymerization of the MT from the tip. The depolymerization rate is, in general, length
and motor-concentration dependent; in addition, MCAK and Kip3p are found to
have vastly different residence times on the MT. These results indicate that
depolymerizing kinesins could be used by the cell for precise length
regulation of microtubules.

The central question of interest for us here is: what is the nature of the
length-distribution of a set of microtubules in steady state, in a solution of
free tubulin and depolymerizing motor proteins (henceforth referred to as {\it
  depolymerizers} for brevity)? In particular, it is
important to understand how the depolymerizers contribute to the formation of
a highly ordered structure such as the metaphase spindle, where most of the MT
would need to be close to the spindle length. This aspect becomes significant,
if we recall that the steady state length distribution of a set of
microtubules undergoing dynamic instability under standard {\it in vitro}
conditions has an exponentially decaying form, without
any preferred length~\cite{dogterom93}. 
%It is also of direct experimental
%interest to understand how the depolymerizing action of different motors is
%related to the specific details of their interaction with the MT,
%eg. transport mechanism and processivity. 

In this letter, we present a theoretical framework for understanding the depolymerizing
activity of a generic motor protein as described above. 
%The model is strongly
%based on published experimental data. 
%and assumptions, wherever used, are clearly
%stated and their validity verified. 
After deriving expressions for the
concentration profile of bound motors on a MT, we calculate the rate of
absorption of motors at the MT tips. We then construct the rate equations for
the length distribution(s) of MT, which are then solved perturbatively in the
limits of low and high motor concentrations. We also show that the model can be
used to analyze available experimental data on motor-induced
depolymerization. 

%%%%%%%%%%%%%%%%%%%%%%%%%%%%%%%%%%%%%%%%%%%%%%%%%%%%%%%%%%%%%%
\begin{figure}
%\onefigure{epl-template.eps}
\begin{center}
\includegraphics[width=0.9\columnwidth]{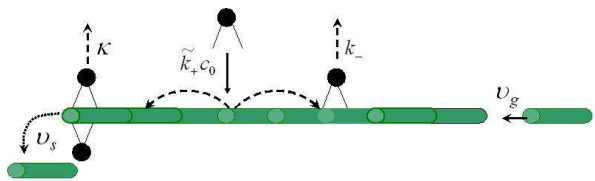}
\end{center}
\vspace{-2.0cm}
  \caption{
An illustration of our model, showing the important kinetic
processes. Depolymerizer motor proteins attach to the surface of 
a MT and are transported along the MT by diffusion and/or directed 
walk. The motors are trapped at the MT tips, where they induce 
depolymerization of the MT. The MT grows at a motor-free tip by 
adding tubulin subunits. Bound motors may detach from the surface 
or the tips of the MT.  
}
  \label{fig0}
\end{figure}
%%%%%%%%%%%%%%%%%%%%%%%%%%%%%%%%%%%%%%%%%%%%%%%%%%%%%%%%%%%%%%

\section{Model description}

Since we are interested in length-related properties, we consider here MT of
finite length (unlike earlier theoretical approaches ~\cite{klein05,howard06}
to the problem, which concentrated  on semi-infinite MT). 
In our model, a depolymerizer binds to a
MT  at a rate ${\tilde k}_{+}c_0$ per unit length ($c_0$ being the depolymerizer
concentration in solution), and detaches at a rate $k_{-}$. A bound 
generic depolymerizer
undergoes a biased random walk with drift speed $v_0$ and diffusion coefficient
$D$. Once a depolymerizer reaches a tip of the MT, it depolymerizes
the filament at a rate $v_s$ (measured in $\mu$m min$^{-1}$) 
while remaining attached to the tip. The depolymerizers accumulate at the tip,
and detach from there at a rate ${\kappa}$, which characterizes its 
{\it processivity}(the tendency to stay attached to 
the MT tip while depolymerizing it). The tips, therefore,
act as sinks for the depolymerizers which could accumulate there. The net rate
of depolymerization, $v_d$, is a stochastic quantity, 
and depends on the number of accumulated motors. We 
further assume that a GTP cap is always present at a MT tip so that  
a depolymerizer-free tip will grow at a rate $v_g$ by adding tubulin
sub-units (i.e., GTP hydrolysis is assumed to be sufficiently slow). 
An illustration of the model is provided in Fig.~\ref{fig0}.

The MT-bound depolymerizer density profile will be denoted by $c(z,t)$ with $0\leq z\leq
\ell$, $\ell$ being the length of the MT at time $t$. The kinetics of $c(z,t)$,
in a frame of reference attached to the moving MT tip is
given by the equation,

\begin{equation}
\frac{\partial c}{\partial t}=D\frac{\partial^2 c}{\partial
  z^2}-v\frac{\partial c}{\partial z}-k_{-}c+{\tilde k}_{+}c_0,
\label{eq1}
\end{equation}
where $v=v_0-v_m$ and $v_m=v_g$ for a growing MT and $v_m=-v_d$ for a
shrinking MT. Note that steric exclusion effects between motors have
been neglected here for simplicity. In general, $v_0$ depends on $c$; 
however, at sufficiently small $c$, modification of the last term of 
eq.(\ref{eq1}) to incorporate steric exclusion is equivalent to addition 
of just a constant to the detachment rate $k_{-}$\cite{klein05}. 

So far as the surface transport on the MT is concerned, the MT 
tips act as ``traps'' for the depolymerases; however, a depolymerase can 
escape from the ``trap'' only by detachment from the MT during its 
MT-depolymerizing activity. Therefore, we treat the kinetics of the
tip-absorbed motor population separately from the MT surface-bound (mobile)
density of motors. The boundary conditions on $c(z,t)$, which are needed 
to solve eq.~(\ref{eq1}), should reflect the ``trapping'' of motors at
the tips, and the simplest case is absorbing boundary conditions: 
$c(0,t)=c(\ell,t)=0$ for all $t$. Using time-independent boundary conditions 
requires that length fluctuations due to growth/shrinkage are sufficiently 
small, and the conditions for the same are derived now.
We consider the $v_0=0$ case first. A single motor 
typically spends a residence time
$\tau_r\sim 1/k_{-}$ on a sufficiently long MT ($\ell\gg
\sqrt{D\tau_r}$), or otherwise gets absorbed at a tip within a shorter time interval
$\sim \ell^2/D$. Clearly, only those motors that bind to the MT within a `depletion zone'
of length $\ell_d\sim \sqrt{D\tau_r}$ from a tip get absorbed at the tip, within a time
interval $\sim \ell_d^2/D\sim \tau_r$, the rest will get detached before they can
reach the tip. The length change over $\tau_r$ is 
$\delta \ell\sim v_m\tau_r$ with $v_m=v_g$ or $-v_s$, and the condition 
$|\delta \ell|\ll \ell$ is satisfied by $|v_m|\ll\sqrt{Dk_{-}}$. In the case of a
motor which undergoes directed walk with velocity 
$v_0$, the corresponding condition turns out to be $|v_m|\ll v_0$. 

%%%%%%%%%%%%%%%%%%%%%%%%%%%%%%%%%%%%%%%%%%%%%%%%%%%%%%%%%%%%%%%%%%%
%\begin{table}
%\caption{Table caption.}
%\label{tab.1}
%\begin{center}
%\begin{tabular}{lcr}
%first  & table & row\\
%second & table & row
%\end{tabular}
%\end{center}
%\end{table}

\begin{table}
\caption{A list of experimental parameter values for MCAK~\cite{howard06} and
    Kip3p ~\cite{varga06} depolymerizers. The depolymerization rate is a
    function of length and motor concentration.}
 \label{tab:1}
\begin{center}
  \begin{tabular}{lcr}
%\hline
\hline
Quantity  & MCAK & Kip3p\\
\hline 
$D$ & 0.38 $\mu$m$^2$s$^{-1}$ & - \\
%\hline
$v_0$ & - & 3.6 $\mu$m min$^{-1}$\\
%\hline
%\hline
$v_d$ & $<4 \mu$m min$^{-1}$ &
$< 2\mu$m min$^{-1}$\\
%\hline
${\tilde k}_+$ & 0.64nM$^{-1}\mu$m$^{-1}$s$^{-1}$& assumed same\\
$k_{-}$ & 1.21 s$^{-1}$ & $\sim 0.004$s$^{-1}$ \\
$\kappa$ & 0.5s$^{-1}$ & 0.03 s$^{-1}$\\ 
\hline
%\hline
  \end{tabular}
 \end{center}
\end{table}
%%%%%%%%%%%%%%%%%%%%%%%%%%%%%%%%%%%%%%%%%%%%%%%%%%%%%%%%%%%%%%%%%%%
Table 1 lists the various experimentally measured parameters for the two
depolymerizers MCAK and Kip3p. For the purely diffusing MCAK, 
the above condition is satisfied for $|v_m|\ll 27\mu$m min$^{-1}$, and the analysis
based on fixed boundary conditions should work well for growth rates of physiological interest. 
For walking Kip3p, the condition is marginally satisfied for low growth/shrinkage rates of $|v_m|\ll
  1\mu$m min$^{-1}$. Nevertheless, we carry our analysis for the general case
  of a depolymerizer with directed motion as well as diffusion, assuming that the
  condition of small length fluctuations is satisfied. In particular,
  we will henceforth assume that $v\simeq v_0$ in eq.~(\ref{eq1}). 

\section{Motor density profile and absorption rates}

The steady-state density profile $c(z)$ of the motors, obtained from 
eq.~(\ref{eq1}), under the absorbing boundary conditions is
\begin{eqnarray}
c(z)=K\bigg[1-\frac{1}{\sinh(\beta
    v\ell/2D)}\bigg(e^{-\frac{v(\ell-z)}{2D}}\sinh(\beta vz/2D)+\nonumber\\
e^{\frac{vz}{2D}}\sinh(\beta v(\ell-z)/2D)\bigg)\bigg]
\label{eq2}
\end{eqnarray}

where $\beta=\sqrt{1+4Dk_{-}/v^2}$ and $K={\tilde k}_{+}c_0/k_{-}$. 
We emphasize that the function $c(z)$, being the density profile of 
only the motors not {\it absorbed} at the MT tips, vanish at $z=0$ and $z=\ell$. 
However, the total concentration of the depolymerizers at the MT tips 
is non-vanishing because of the accumulation of the absorbed depolymerizers 
there. We also note that eq.~(\ref{eq2}) is invariant under the combined
transformations $v\to -v$ and $z\to \ell-z$. 

It is interesting to look at the limiting behavior of the density profile in
eq.~(\ref{eq2}) in the limits $v\to 0,D>0$ (pure diffusion) and $D\to 0,v>0$
(pure walk), since these special cases are of experimental interest. In the
first case, the distribution has the following well-defined form, which is
symmetric about $z=\ell/2$:

\begin{eqnarray}
c_d(z)\equiv \lim_{v\to
  0}c(z)=K\bigg[1-\frac{1}{\sinh(\ell/\lambda_d)}\bigg(\sinh(z/\lambda_d)+\nonumber\\
\sinh((\ell-z)/\lambda_d)\bigg)\bigg]~~~;\lambda_d=\sqrt{D/k_{-}}
\label{eq3}
\end{eqnarray}

In the second case, after taking the limit $D\to 0$, we find that 

\begin{equation}
c_v(z)\equiv \lim_{D\to 0} c(z)=K[1-\exp(-z/\lambda_w)]~~~~;\lambda_w=v/k_{-}
\label{eq4}
\end{equation}

Note that eq.~(\ref{eq4}) is non-vanishing as
$z\to \ell$, as expected for a pure directed walk towards the plus end. This does not contradict
the imposed boundary conditions, because we still have $\lim_{D\to
  0}\lim_{z\to l}c(z)=0$. 

Since the kinetics at the minus end of a MT is typically much slower than that
at the plus end, we concentrate only on the plus-end kinetics for the rest of this paper, 
though motor accumulation is allowed to occur at both ends.
The rate of absorption of motors at the plus-end is given by 
$\nu(\ell)=-D(\partial c/\partial z)_{z=\ell}$, and has the general form:

\begin{eqnarray}
\nu(\ell)=\frac{vK}{2\sinh(\beta v\ell/2D)}\bigg[\beta\bigg(\cosh(\beta
v\ell/2d)-e^{v\ell/2D}\bigg)+\nonumber\\ \sinh(\beta v\ell/2D)\bigg].
\label{eq5}
\end{eqnarray}

$\nu(\ell)$ vanishes at $\ell=0$, and is a monotonically increasing function of
$\ell$ for all values of $v$ and $D$. The saturation value for the general
$v>0, D>0$ case is given by

\begin{equation}
\nu_{\max}=vK(1+\beta)/2.
\label{eq6}
\end{equation}

For pure diffusion and pure walk, eq.~(\ref{eq5}) reduces, respectively, to the limiting 
forms,

\begin{equation}
\nu_d(\ell)={\tilde k}_{+}c_{0}\lambda_{d}\tanh(\ell/2\lambda_{d})~~~~~;v\to 0
\label{eq7}
\end{equation}

and 

\begin{equation}
\nu_v(\ell)={\tilde
  k}_{+}c_{0}\lambda_{w}\left[1-\exp(-\ell/\lambda_{w})\right]=vc_v(l)~~~~;D\to 0.
\label{eq8}
\end{equation}

For short MT ($\ell\ll \lambda_d$ or $\lambda_w$ 
respectively), $\nu(\ell)\approx \frac{1}{2}{\tilde k}_{+}\ell$ in the first 
case and $\nu(\ell)\approx {\tilde k}_{+}\ell$ in the second; the difference 
of the factor of 2 arises from the fact that a diffusing depolymerizer 
can target either of the tips. In the opposite limit of long MT, both 
the rates approach their saturation values ${\tilde k}_{+}c_0\lambda_{d}$ 
and ${\tilde k}_{+}c_0\lambda_{w}$, respectively.
(which may also be derived directly from eq.~(\ref{eq6})).

\section{Rate Equations}

We will now study how a given depolymerizer would affect the length
distribution of a set of MT, when a steady state is reached by a
balance between depolymerizer binding/detachment and MT
polymerization/depolymerization processes. For simplicity, we neglect the three-dimensional
structure of the MT filament and imagine the MT as a linear polymer, made of
sub-units of length $b$. We denote by $P_n(m,t)$ the fraction of polymers with
$m$ sub-units and $n$ absorbed depolymerizers at the plus-end. Let $p_g=v_g/b$ be the
probability per unit time for attachment of a subunit to a free tip 
and $p_s=v_s/b$ be the rate of removal of subunits per motor (we assume
henceforth that the rate of sub-unit removal increases linearly with the number of motors). 
The rate equations for $P_n(m,t)$ are as follows: For $m = 1$, 

\begin{eqnarray}
\frac{\partial P_0(1)}{\partial t}=-p_g P_0(1)+\kappa P_1(1),~~~~~~~~~~~~~~~~~~~~~~~~~\nonumber\\
\frac{\partial P_1(1)}{\partial t}=p_s P_1(2)-\kappa P_1(1)+\nu(b)P_0(1),~~~~~~~
\label{eq9}
\end{eqnarray}
while for $m\geq 2$ and $n\geq 1$,

\begin{eqnarray}
\frac{\partial P_0(m)}{\partial t}=-p_g [P_0(m)-P_0(m-1)]+\nonumber\\\kappa
P_1(m)-\nu(mb)P_0(m)\nonumber\\
\frac{\partial P_n(m)}{\partial t}=np_s[P_n(m+1)-P_n(m)]+\nonumber\\
(n+1)\kappa
P_{n+1}(m)+\nu(mb)P_{n-1}(m)-\nonumber\\
(n\kappa+\nu(mb))P_n(m)~~~~~~~~
\label{eq10}
\end{eqnarray}

We now focus on the steady state of the model. Since a general analytic
solution is difficult, we will adopt a perturbative approach, and
look at the solutions in the limit of
low and high depolymerizer concentrations. The transition between these regimes
is controlled by the dimensionless ratio $\eta=\nu_{\max}/\kappa$ 
which may be used as the small parameter in the low density perturbation
expansion. This is seen through the following argument. 

Let us define a distribution of the number of accumulated motors: 
$\chi_n\equiv \sum_{m=1}^{\infty}P_n(m)$. It is convenient to
use the numbers $\alpha_n<1$, defined through the relation
$\sum_{m=1}^{\infty}\nu(mb)P_n(m)\equiv \nu_{\max}\alpha_n\chi_n$. From
eq.(~\ref{eq9}) and eq.(~\ref{eq10}), it can be shown that, in steady state, 
$\chi_n$ follows a Poisson-like distribution of the form 
$\chi_n=\chi_0(\eta^n/n!)\prod_{i=0}^{n-1}\alpha_i$ for $n\geq 1$,
where $\chi_0$ is determined through normalization: $\sum_{n=0}^{\infty}\chi_n=1$.
%In the limit $\eta\ll 1$, $\chi_n\sim O(\eta^n)$, and therefore the different
%$P_n(m)$ are $O(\eta^n)$ or smaller. 

{\it Low motor concentration:} For $\eta\ll 1$, we can neglect $P_n$ with $n\geq 2$ in eq.~(\ref{eq10}), 
in comparison with $P_0$ and $P_1$. 
%Let us further simplify eq.(\ref{eq9}) by studying the
%continuum limit $m\gg 1$ in terms of the
%length $l\approx mb$. 
By keeping terms upto $O(\eta)$, in the continuum limit ($m\gg 1, \ell\approx mb$), we arrive 
at a single combined equation for $P_0$ and $P_1$ in steady state:

\begin{equation}
v_g\frac{\partial P_0}{\partial \ell}=-\nu(\ell)P_0(\ell)+\kappa
P_1(\ell)=v_s\frac{\partial P_1}{\partial \ell}
\label{eq11}
\end{equation}

which has the general solution (for $\ell\gg b$)

\begin{equation}
P_0(\ell)=C\exp\bigg(\frac{\kappa}{v_s}\ell-\frac{1}{v_g}\int_{0}^{\ell}\nu(\ell^{\prime})d\ell^{\prime}\bigg)=\frac{v_s}{v_g}P_1(\ell)
\label{eq12}
\end{equation}

where $C$ is a constant of normalization. In the absence of a boundary, this solution is normalizable only 
when $\eta>\eta_c$, where $\eta_c=v_g/v_s $ 
implicitly gives a critical motor concentration, below which a well-defined steady state distribution does
not exist. The above result for $\eta_c$ is, however, strictly true only when $v_g/v_s\ll 1$ since the
present analysis assumes $\eta\ll 1$. In the more general case, a steady state
should still exist at sufficiently large $\eta$, but the critical
concentration would be given by a relation of the form 
$\eta_c=f(v_g/v_s)$, with $f(x)\simeq x$ as $x\to 0$. 

{\it High motor concentration:} To analyze the case $\eta\gg 1$, 
it is convenient to re-express eq.~(\ref{eq9}) and
eq.~(\ref{eq10}) (in the continuum limit) as

\begin{eqnarray}
\frac{v_g}{\nu_{\max}}\frac{\partial P_0}{\partial
  \ell}=\eta^{-1}P_1(\ell)-f(\ell)P_0(\ell)\nonumber\\
-\frac{nv_s}{\nu_{\max}}\frac{\partial P_n}{\partial
  \ell}=(n+1)\eta^{-1}P_{n+1}(\ell)+f(l)P_{n-1}(\ell)-\nonumber\\
(n\eta^{-1}+f(\ell))P_n(\ell),
\label{eq13}
\end{eqnarray}

where the function $f(\ell)\equiv \nu(\ell)/\nu_{\max}\leq 1$. 
We now expand all $P_n(l)$ in a perturbation series in $\eta^{-1}$: 
$P_n(\ell)=\sum_{m=0}^{\infty}\eta^{-m}P_n^{(m)}(\ell)$ for $n\geq 0$, and then
substitute into eq.~(\ref{eq13}). The zeroth order distributions are then
found to satisfy the iterative equation,

\begin{equation}
-\frac{nv_s}{\nu_{\max}}\frac{\partial P_n^{(0)}}{\partial \ell}=f(\ell)[P_{n-1}^{(0)}(\ell)-P_n^{(0)}(\ell)]
\label{eq14}
\end{equation}

and $P_0^{(0)}(\ell)=P_0^{(0)}(0)\exp(-v_g^{-1}\int_0^{\ell}d{\ell}^{\prime}
\nu(\ell^{\prime}))$ from eq.~(\ref{eq13}). The solution of eq.~(\ref{eq14})
is found after a few elementary calculations:

\begin{equation}
P_n^{(0)}(\ell)=P_n^{(0)}(0)e^{-\frac{1}{v_g}\int_{0}^{\ell}d\ell^{\prime}\nu(\ell^{\prime})}.
\label{eq15}
\end{equation}

where 

\begin{equation}
P_n^{(0)}(0)=P_0^{(0)}(0)\prod_{m=1}^{n}\frac{v_g}{v_g+mv_s}.
\label{eq16}
\end{equation}

The unknown constant $P_0^{(0)}$ is determined by normalization. 
Eq.~(\ref{eq15}) is a monotonically decreasing function of
$\ell$, however, non-monotonic terms appear in the first order and above.

%%%%%%%%%%%%%%%%%%%%%%%%%%%%%%%%%%%%%%%%%%%%%%%%%%%%%%%%%%%%%%
\begin{figure}
%\onefigure{epl-template.eps}
\begin{center}
\psfrag{'fort.12'}{$\zeta=1.2$}
\psfrag{'fort.16'}{$\zeta=1.6$}
\psfrag{'fort.20'}{$\zeta=2.0$}
\psfrag{P(x)}{$P_0(x)$}
\psfrag{<x>}{$\langle x\rangle$}
\psfrag{x}{$x$}
\psfrag{c}{$\zeta$}
\includegraphics[width=0.7\columnwidth]{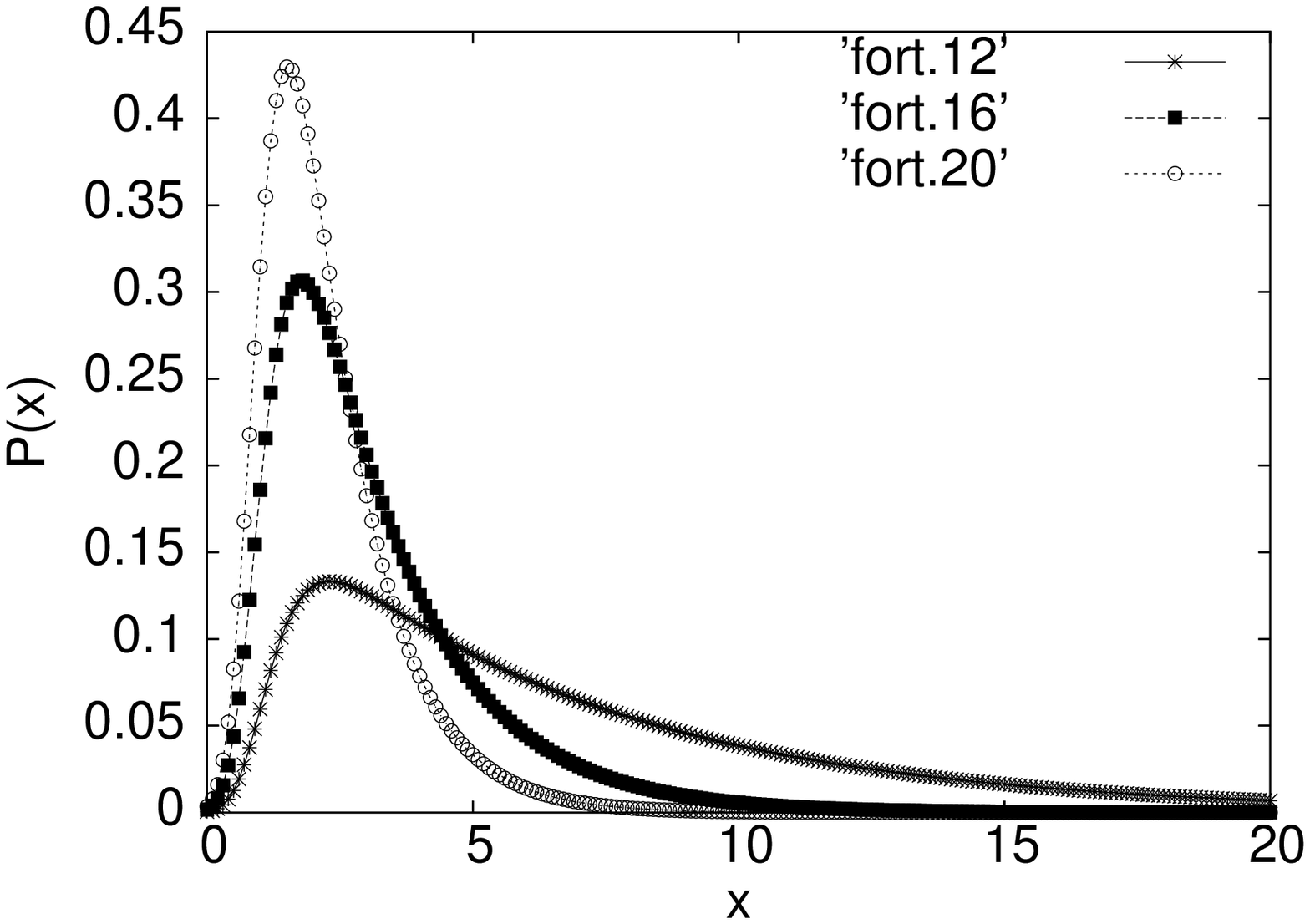}
\includegraphics[width=0.4\columnwidth]{FIG2inset.ps}
\end{center}
\vspace{-9.5cm}
 \caption
{The distribution in eq.~(\ref{eq17}) (normalized such that $\int P_0(\ell)d\ell=1$) is plotted against
  the dimensionless `length' $x=\ell/2\lambda_d$(eq.~(\ref{eq3})) 
for three different values of $\zeta=\eta/\eta_c$.
The other parameters were fixed at the MCAK values (table 1). We assumed
  $v_s\simeq v_d$ and $v_g\simeq 1\mu$m min$^{-1}$. From table 1, 
$\lambda_d\simeq 0.56\mu$m for MCAK. Inset: the mean length as a function of
  $\zeta$ (computed numerically, $\delta x=10^{-5}$ everywhere). Using the
  parameters in table 1, the critical
  MCAK concentration corresponding to $\eta_c$ is $~ 1.39v_g/v_s$ nM. 
}
  \label{fig1}
\end{figure}
%%%%%%%%%%%%%%%%%%%%%%%%%%%%%%%%%%%%%%%%%%%%%%%%%%%%%%%%%%%%%%

%%%%%%%%%%%%%%%%%%%%%%%%%%%%%%%%%%%%%%%%%%%%%%%%%%%%%%%%%%%%%%
\begin{figure}
%\onefigure{epl-template.eps}
\begin{center}
\psfrag{'fort.12'}{$\zeta=1.2$}
\psfrag{'fort.16'}{$\zeta=1.6$}
\psfrag{'fort.20'}{$\zeta=2.0$}
\psfrag{P(x)}{$P_0(x)$}
\psfrag{<x>}{$\langle x\rangle$}
\psfrag{x}{$x$}
\psfrag{c}{$\zeta$}
\includegraphics[width=0.7\columnwidth]{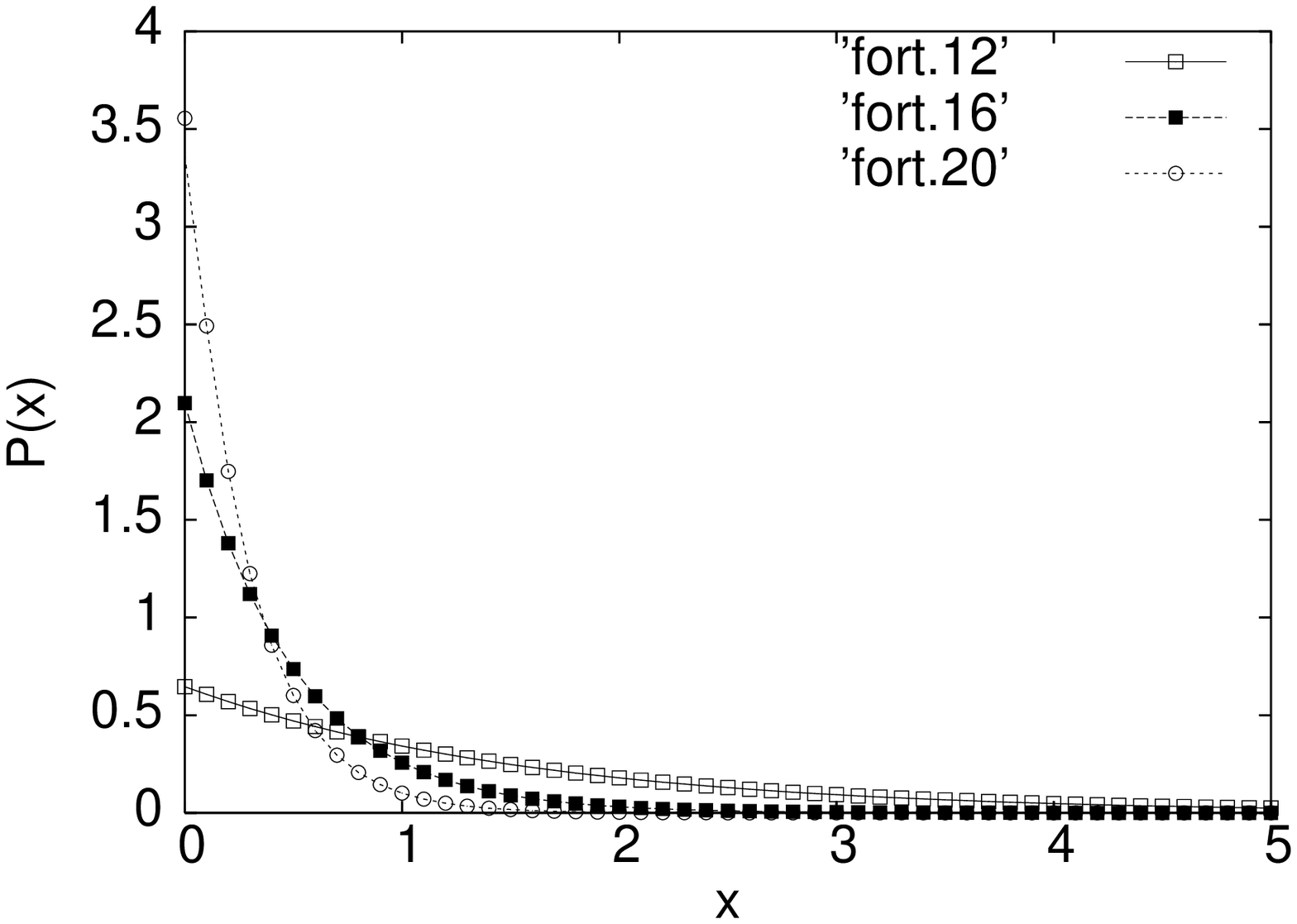}
\includegraphics[width=0.4\columnwidth]{FIG3inset.ps}
\end{center}
\vspace{-10.0 cm}
  \caption{Similar to the previous figure, using the distribution in
    eq.(\ref{eq18}) and Kip3p parameter values (table 1). Here, $x=\ell/\lambda_w$ (eq.(\ref{eq4}))with 
$\lambda_w\simeq 15\mu$m for Kip3p. The critical Kip3p concentration
corresponding to $\eta_c$ is $\sim 0.003 v_g/v_s$ nM.
}
  \label{fig2}
\end{figure}
%%%%%%%%%%%%%%%%%%%%%%%%%%%%%%%%%%%%%%%%%%%%%%%%%%%%%%%%%%%%%%

\section{Application to specific cases}

Our results so far have been general, within the limits of validity of the
assumptions stated in the second section. We will now examine their
implications for the specific depolymerizer motor proteins studied in experiments.

The solution obtained in eq.~(\ref{eq12}) in the limit of low concentrations 
is a non-monotonic function of the
length, increasing exponentially as $\sim e^{(\kappa/v_s)\ell}$ for small $\ell$, and
decreasing exponentially at large $\ell$, with a peak at an intermediate
value. The location of this peak depends on the ratio $\kappa/v_s$, and for
fixed $v_s$, the peak is more pronounced for large $\kappa$ (low processivity)
and vice-versa. The depolymerizers MCAK and Kip3p have very
different processivities (table 1), and it is therefore interesting to look at
these specific cases in more detail. Eq.~(\ref{eq12}) reduces to the following 
forms in the pure diffusion ($v\to 0, D>0$) and pure walk ($v>0, D\to 0$) limits
respectively:

\begin{equation}
P_0(\ell)\propto e^{\kappa \ell/v_s}
\left[\cosh(\ell/2\lambda_d)\right]^{-\frac {2K\lambda_d^2}{v_g}}~ (v_0=0)
\label{eq17}
\end{equation}

\begin{equation}
P_0(\ell)\propto \exp\left[\bigg(\frac{\kappa}{v_s}-\frac{K\lambda_w}{v_g}\bigg)\ell-\frac{K\lambda_w^2}{v_g}e^{-\ell/\lambda_w}\right](D=0)
\label{eq18}
\end{equation}

%%%%%%%%%%%%%%%%%%%%%%%%%%%%%%%%%%%%%%%%%%%%%%%%%%%%%%%%%%%%%%%%%%
\begin{figure}
%\onefigure{epl-template.eps}
\begin{center}
\psfrag{fluct}{$\frac{\Delta x}{\langle x\rangle}$}
\psfrag{c}{$\zeta$}
\includegraphics[width=0.6\columnwidth]{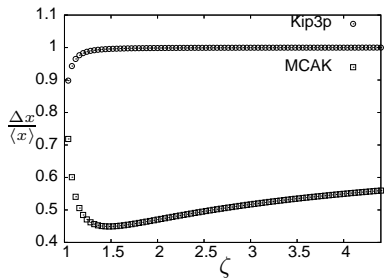}
\end{center}
\vspace{-4.0cm}
  \caption{The ratio of the standard deviation to the mean length of the
    distributions in fig.\ref{fig1} and fig.\ref{fig2} as a
    function of  $\zeta=\eta/\eta_c$.}
  \label{fig3}
\end{figure}
%%%%%%%%%%%%%%%%%%%%%%%%%%%%%%%%%%%%%%%%%%%%%%%%%%%%%%%%%%%%%%%%%%

Fig.~\ref{fig1} and fig.~\ref{fig2} show the normalized forms of the distributions in
eq.~(\ref{eq17}) and eq.~(\ref{eq18}) respectively, plotted against the
dimensionless length variables $x=\ell/\lambda_d$ and $x=\ell/\lambda_w$, for
three values of $\zeta=\eta/\eta_c >1$ (the empirical values of 
$\kappa, D, v_0$ and $v_s$ are taken from Table 1). 
Although, as we explained in the beginning, our
theory is more applicable to MCAK than Kip3p, the difference between the two
cases is nevertheless striking: the MCAK distribution shows a peak, while the Kip3p
distribution is essentially a monotonically decreasing function (the peak is too
close to the origin to be visible in the plot). The contrasting behavior is primarily due to their very different processivities.

In order to characterize the MCAK and Kip3p-induced distributions better, we also looked at 
how the lengths are spread about the mean value
in each of these cases. In fig.~\ref{fig3}, we plot the relative MT length fluctuation $\Delta x/\langle
x\rangle$ for MCAK and Kip3p, as a function of the 
dimensionless ratio $\zeta=\eta/\eta_c$, where $\Delta x=\sqrt{\langle
  x^2\rangle-\langle x\rangle^2}$ is the standard deviation. The relative
fluctuation for MCAK is typically smaller than 1, and, interestingly, also
shows a minimum of $\sim 0.3$ at $\zeta\simeq 1.37$. For Kip3p, on the other
hand, the relative fluctuation increases rapidly with $\zeta$ and saturates at
unity, which is characteristic of a purely exponential distribution. 
MCAK appears to produce a tighter control of length than Kip3p because of its lower processivity.

\section{Length-dependent depolymerization}

%%%%%%%%%%%%%%%%%%%%%%%%%%%%%%%%%%%%%%%%%%%%%%%%%%%%%%%%%%%%%%
\begin{figure}
%\onefigure{epl-template.eps}
\begin{center}
\psfrag{'fort.12'}{$\zeta=1.2$}
\psfrag{'fort.16'}{$\zeta=1.6$}
\psfrag{'fort.20'}{$\zeta=2.0$}
\psfrag{n}{$\langle n\rangle_{\ell}$}
\psfrag{l}{$\ell$($\mu$m)}
\psfrag{c}{$c_0$(nM)}
\psfrag{'conc1'}{1nM}
\psfrag{'fort.15'}{4nM}
\psfrag{'fort.16'}{$\ell$=5$\mu$m}
\includegraphics[width=0.7\columnwidth]{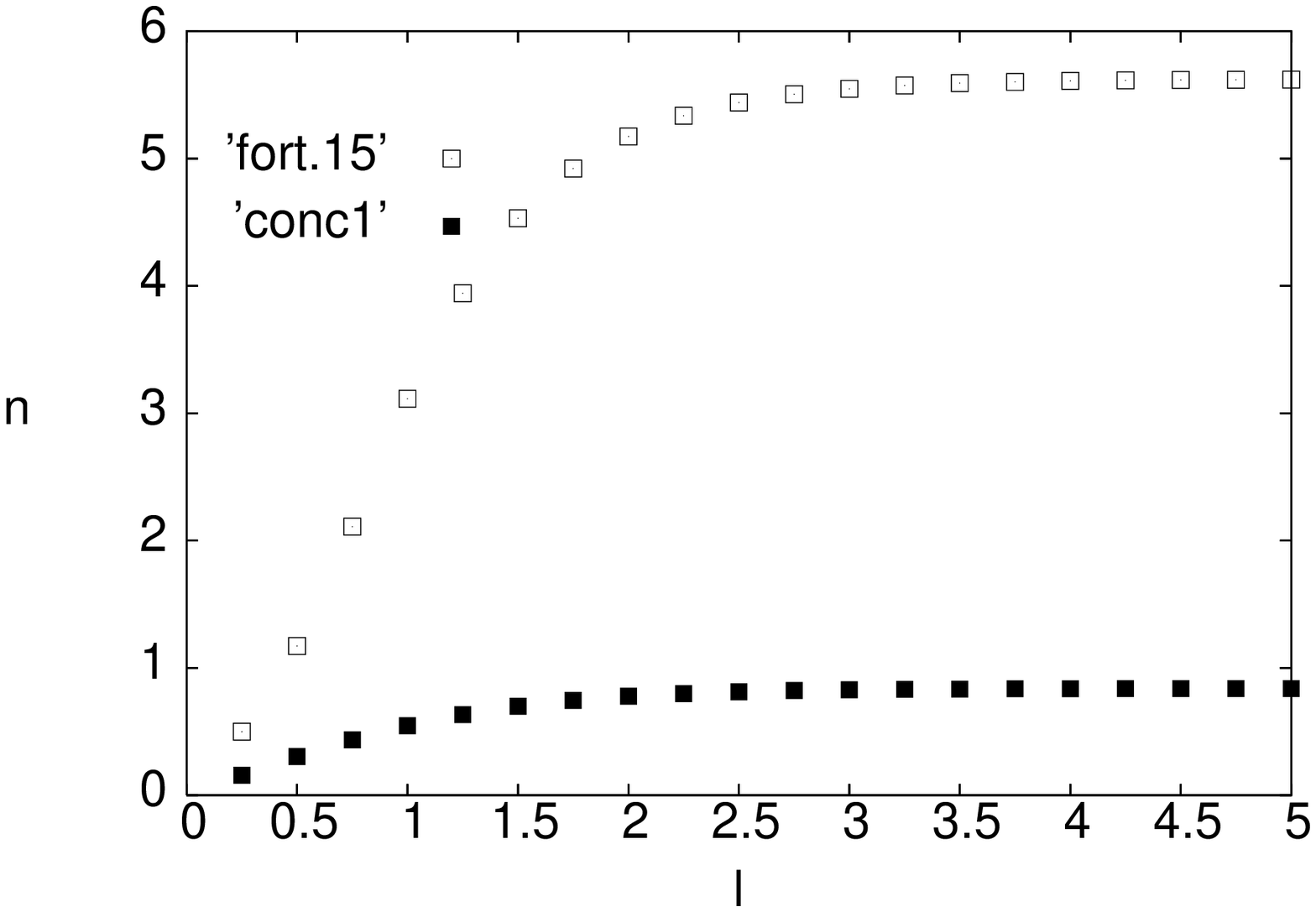}
\includegraphics[width=0.4\columnwidth]{FIG5inset.ps}
\end{center}
\vspace{-9.7cm}
  \caption{The figure shows the mean number of tip-absorbed motors (computed
    numerically using eq.~\ref{eq21}), plotted as
    a function of the MT length for MCAK, for two values of the bulk motor
    concentration $c_0$. The inset shows the saturation value
    (computed at $\ell=5\mu$m) plotted against $c_0$.
}
  \label{fig5}
\end{figure}
%%%%%%%%%%%%%%%%%%%%%%%%%%%%%%%%%%%%%%%%%%%%%%%%%%%%%%%%%%%%%%
We will now focus on the analysis of the existing experimental results on motor-induced
depolymerization within our model. The rate of depolymerization of a
microtubule depends on the number $n$ of motor
proteins accumulated at the tip. Experiments with fixed microtubule length measure the
mean depolymerization rate $\langle v_d\rangle=v_s\langle n\rangle_\ell$, where $\langle
n\rangle_{\ell}$ gives the mean number of motors attached to a microtubule of
length $\ell$. To calculate this average, we define the distribution $Q_\ell(n)$ for the number of
motors attached to the tip of a MT of length $\ell$. We assume that a motor
protein initially bound to one protofilament will continue to travel along it
until it reaches a tip, without hopping between filaments. 
In this case, absorption at a certain protofilament tip is
possible only if that particular tip is free, the probability of which is given
by $(1-n/n^*)$, where $n$ is the total number of motors attached at the MT tip
and $n^*$ is the maximum number of motors that can be absorbed at a given
time. Since a MT usually has 13 protofilaments, we assume $n^*=13$ for concreteness.

The rate equations for $Q_{\ell}(n)$ are as follows. For motor-free tips, we have

\begin{eqnarray}
\frac{\partial Q_\ell(0)}{\partial t}=  \kappa Q_{\ell}(1)-\nu(\ell)Q_{\ell}(0)\nonumber\\
\frac{\partial Q_{\ell}(n)}{\partial t}=(n+1)\kappa
Q_{\ell}(n+1)+\nonumber\\
\nu(\ell)\bigg(1-\frac{n}{n^*}\bigg)\bigg[Q_{\ell}(n-1)-Q_{\ell}(n)\bigg]+\nonumber\\
Q_{\ell}(n-1)\frac{\nu (l)}{n^*}-n\kappa Q_{\ell}(n)~~~~;1\leq n< n^*\nonumber\\
\frac{\partial Q_{\ell}(n^*)}{\partial t}=-n^*\kappa
Q_{\ell}(n^*)+\frac{\nu(\ell)}{n^*}Q_{\ell}(n^*-1)
\label{eq21}
\end{eqnarray}

The steady state solution of eq.~(\ref{eq21}) is given by

\begin{equation}
Q_{\ell}(n)=Q_{\ell}(0)\frac{1}{n!}\bigg(\frac{\nu(\ell)}{\kappa}\bigg)^n\prod_{j=1}^{n-1}(1-\frac{j}{n^*})~~~~1\leq
n\leq n^*
\label{eq22}
\end{equation}

where $Q_{\ell}(0)$ is fixed by normalization:
$\sum_{n=1}^{n^*}Q_{\ell}(n)=1$. The mean number of tip-accumulated motors is
then given by $\langle n\rangle_{\ell}=\sum_{n=1}^{n^*}nQ_{\ell}(n)$.

%%%%%%%%%%%%%%%%%%%%%%%%%%%%%%%%%%%%%%%%%%%%%%%%%%%%%%%%%%%%%%
\begin{figure}
%\onefigure{epl-template.eps}
\begin{center}
\psfrag{n}{$\langle n\rangle_{\ell}$}
\psfrag{l}{$\ell$($\mu$m)}
\includegraphics[width=0.7\columnwidth]{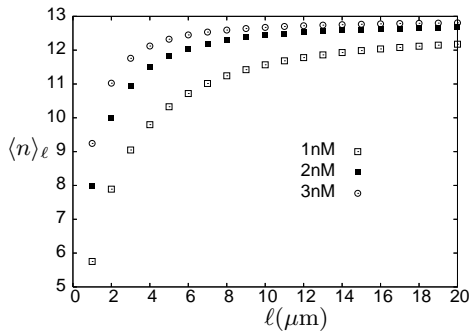}
\end{center}
\vspace{-4.7cm}
  \caption{The average number of tip-absorbed motors in the presence of Kip3p,
    for three different values of $c_0$. Note the difference in length scales
    between this and the previous figure.  
}
  \label{fig6}
\end{figure}
%%%%%%%%%%%%%%%%%%%%%%%%%%%%%%%%%%%%%%%%%%%%%%%%%%%%%%%%%%%%%%

In fig.~\ref{fig5}, we have plotted numerically computed $\langle n\rangle_{\ell}$ as a function of
$\ell$ in the case of MCAK for two different motor concentrations $c_0$. It is found
that appreciable length-dependence of $\langle n\rangle_{\ell}$, and hence the
mean depolymerization velocity $\langle v_d\rangle$, is found only for $\ell<
2\mu$m. This is in agreement with experiments, where no length-dependence of
depolymerization was observed in the case of MCAK, for MT longer than
2$\mu$m. The saturation value of the depolymerization rate (plotted
in the inset, for $\ell=5\mu$m) depends strongly on the motor
concentration, and this curve also agrees well with experimental
results~\cite{howard06} and previous theoretical predictions~\cite{klein05}.

We now turn to the case of Kip3p, in which case a strong length-dependence of
depolymerization was observed in experiments~\cite{varga06}. Fig.~\ref{fig6}
shows the theoretical plot of $\langle n\rangle_{\ell}$ against $\ell$ for
three motor concentrations. The length-dependence here is much stronger than
MCAK, and for low concentrations, saturation is not reached even for
$\ell=20\mu$m. These observations qualitatively agree with the
corresponding experimental results. However, a close quantitative agreement
(in particular, experiments show a sharp rise in depolymerization rate between $c_0$=3.3nM
and $c_0$=5.8nM) is missing in this case. 

\section{Conclusions}

To conclude, in this letter, we have formulated a general theory for 
the action of MT-depolymerizing motor proteins. Specifically, we 
considered the limit where the motor density profile becomes stationary  
much before the MT length distribution reaches its steady-state. 
We showed that the rate of accumulation of
the motors at the MT tips, and consequently their depolymerizing activity
itself, is strongly length-dependent. This has a rather pronounced effect on the length
distribution of the MT, which displays a peak before decaying exponentially
at large $\ell$. Interestingly, the processivity of the depolymerizer plays an important role in
determining the nature of the length distribution: a depolymerizer with low processivity 
produces a more pronounced peak in the length distribution, and is likely to be more useful for precise
length regulation. 

Our theory is relevant to future experimental studies
on the role of depolymerizing motor proteins in length regulation of
MT, especially in the context of formation of the metaphase spindle. 
Indeed, very recent experiments using fluroscent speckle microscopy
have shown that the length distribution of individual MT in a meiotic spindle
is strongly non-monotonic~\cite{yang07}. It would be interesting if the
predictions made in this paper could be put to test in future {\it in vitro}
experiments with depolymerizing motor proteins, where the MT lengths are carefully monitored.

%See fig.~\ref{fig.1}, table~\ref{tab.1} and eq.~(\ref{eq.1}).
%See also~\cite{b.a,b.b}.

%\begin{equation}
%\label{eq.1}
%0\neq1
%\end{equation}

%\begin{figure}
%\onefigure{epl-template.eps}
%\caption{Figure caption.}
%\label{fig.1}
%\end{figure}

%\begin{table}
%\caption{Table caption.}
%\label{tab.1}
%\begin{center}
%\begin{tabular}{lcr}
%first  & table & row\\
%second & table & row
%\end{tabular}
%\end{center}
%\end{table}

\acknowledgments

BSG thanks ASICTP (Italy) for hospitality, where part of this work was carried out, 
and acknowledges financial support from DST (India) through a SERC 
Fast-track fellowship. DC thanks F. J\"ulicher and J. Howard for 
useful discussions and acknowledges financial support from CSIR 
(India) and MPI-PKS (Germany).


\begin{thebibliography}{100}

\bibitem{desai97} Desai A. and Mitchison. T.J.,
  Annu. Rev. Cell. Dev. Biol.{\bf 13}, 83 (1997).

\bibitem{hunter00} Hunter A.W. and Wordeman L., J. Cell Sci.{\bf 113}, 4379 (2000).

\bibitem{moore05} Moore A.T. et al, J. Cell Biol. {\bf 169}, 391 (2005).

\bibitem{kinoshita06} Kinoshita K. et al, J. Mus. Res. Cell. Mot.{\bf 27},
    107 (2006).

\bibitem{howard07} Howard J. and Hyman A.A., Curr. Opin. Cell Biol.{\bf 19},
  31 (2007).

\bibitem{wordeman05} Wordeman L., Curr. Opin. Cell Biol.,{\bf 17}, 82 (2005).

\bibitem{hunter03} Hunter A.W. et al, Mol. Cell{\bf 11}, 445 (2003).

\bibitem{howard06} Helenius J. et al, Nature {\bf 441}, 115 (2006).

\bibitem{varga06} Varga V. et al, Nat. Cell Biol {\bf 8}, 957 (2006).


\bibitem{gupta06} Gupta M. L. et al, Nat. Cell Biol.{\bf 8}, 913 (2006).

\bibitem{dogterom93} Dogterom  M. and Leibler S., Phys. Rev. Lett.{\bf 70}, 1347 (1993).


  %\Editor{A. Editor}
  %\Vol{9}
  %\Publ{Publishing house, City}
  %\Year{1939}
  %\Page{666}.



  %\Editor{Editor A.}
  %\Book{Some Book of Interest}
  %\Vol{9}
  %\Publ{Publishing house, City}
  %\Year{1939}
  %\Section{A}.






\bibitem{klein05} Klein  G. A. et al, Phys. Rev. Lett.{\bf 94}, 108012 (2005).


\bibitem{yang07} Yang G. et al, Nat. Cell  Biol.{\bf 9}(11), 1233(2007).
\end{thebibliography}
\end{document}